\documentclass[varenna]{cimento}

\usepackage{graphicx}
\usepackage{physics}
\usepackage{comment}
\usepackage{color}
\usepackage{dcolumn}
\usepackage{bm}
\usepackage{amsmath}
\usepackage{amsfonts,amssymb}
\usepackage{cancel}

\newcommand{\sch}{Schr{\"o}dinger }
\newcommand{\kps}{\ket{\Psi}}
\newcommand{\up}{\uparrow}
\newcommand{\down}{\downarrow}

\renewcommand{\k}{{\bf k}}
\newcommand{\p}{{\bf p}}

\newcommand{\q}{{\bf q}}

\newcommand{\0}{{\bf 0}}
\newcommand{\rvec}{{\mathbf r}}

\newcommand{\ef}{E_F}
\newcommand{\kf}{k_F}
\newcommand{\eb}{\varepsilon_{\rm B}}

\newcommand{\nn}{\nonumber}
\newcommand{\beq}{\begin{equation}}
\newcommand{\eeq}{\end{equation}}

\newcommand{\vol}{\mathcal{V}}

\makeatletter
\renewenvironment{table}[1][\fps@table]
 {%
  \begingroup\edef\x{\endgroup\noexpand\@float{table}[#1]}\x
  \small\@intable@true
  \def\tnote##1{\par$({##1})$}%
 }
 {\@intable@false\end@float}
\makeatother

\title{Fermi polarons and beyond}
\author{M.~M.~Parish \atque J.~Levinsen}

\institute{School of Physics and Astronomy, Monash University, Victoria 3800, Australia}

\shortauthor{M.~M.~Parish \atque J.~Levinsen}

\begin{document}

\maketitle

\begin{abstract}
These lecture notes give a brief introduction to the so-called Fermi-polaron problem, which explores the behaviour of a mobile impurity introduced into an ideal Fermi gas. While this problem has been considered now for more than a decade in ultracold atomic gases, it continues to generate surprises and insights as new quantum mixtures emerge, both in atomic gases and in the solid state. Here we summarise the basic theory for the Fermi polaron with a focus on the three-dimensional case, although the results can be straightforwardly generalised to two dimensions. Our aim is to provide a pedagogical treatment of the subject and we thus cover fundamental topics such as scattering theory and renormalisation. We discuss the ground state of the Fermi polaron and how it is connected to the phase diagram of the spin-imbalanced Fermi gas, and we also give a brief overview of the energy spectrum and non-equilibrium dynamics. Throughout, we highlight how the static and dynamic behaviour of the Fermi polaron is well described using intuitive variational approaches.

\end{abstract}

\newpage 

\tableofcontents

 \section{Introduction}
The problem of a mobile impurity immersed in a quantum medium is ubiquitous in physics, arising in a multitude of systems across a remarkable range of energy scales.
It was first considered in the context of solid-state systems where 
electrons moving through a crystal lattice become modified by interactions with lattice vibrations (phonons), resulting in a new quasiparticle termed a \emph{polaron}~\cite{Landau1933}. Since then, the polaron concept has emerged in a broad variety of systems, 
all the way from proton impurities in neutron stars~\cite{Nemeth1968}---the densest matter in the universe---to atomic impurities in laser-trapped ultracold atomic gases~\cite{Scazza2022}---the most dilute matter in the universe. 
Most recently, it has been realised that the optical response of doped semiconductors can be viewed as a polaron problem~\cite{Sidler2017}, where the optically excited exciton (electron-hole bound state) corresponds to the impurity, and the gas of doped electrons plays the role of the quantum medium.

The focus of these lectures is on the case where the medium is an ideal Fermi gas, a scenario which is readily simulated with cold atoms and which has generated much attention during the past two decades~\cite{Chevy2010,Massignan2014,Levinsen2015,Schmidt2018,Scazza2022}. Despite the background Fermi medium being simple and well understood, this so-called Fermi-polaron problem constitutes a quintessential many-body system that hosts highly non-trivial behaviour. For instance, the impurity can undergo sharp transitions where it abruptly changes its properties such as its mass and statistics. Furthermore, in the case of an infinitely heavy impurity, there is the famous ``orthogonality catastrophe'' first described by Anderson~\cite{Anderson1967}, where the ground state of the impurity interacting with the Fermi sea has \emph{zero} overlap with that of its non-interacting counterpart. Finally, Fermi polarons can provide insight into more complicated many-body systems such as two-component quantum mixtures involving both a finite density of impurity particles and a Fermi gas. In particular, the existence or otherwise of single-impurity transitions will have consequences for the nature of the quantum phase transitions in the quantum mixture.

The investigation of Fermi polarons continues to be an active area of research, providing fundamental insights into quantum many-body phenomena, in part due to the variety of quantum mixtures that keep emerging, both in cold atoms and in semiconductor heterostructures.
In the interest of space, we will restrict ourselves to degenerate atomic Fermi gases, which provide a particularly clean model platform in which to study the quantum impurity problem since the Hamiltonians are generally well understood and tunable, 
without any unwanted disorder.
Moreover, there are now a wide variety of quantum mixtures available with different masses and atomic interactions: 
over the past decade, cold-atom experiments on Fermi polarons~\cite{Schirotzek2009,Nascimbene2009,Kohstall2012,Koschorreck2012,Zhang2012,Wenz2013,Ong2015,Cetina2015,Cetina2016,Scazza2017,Mukherjee2017,Yan2019,Oppong2019,Ness2020,Fritsche2021} have primarily used $^6$Li, $^{40}$K, or mixtures of these atomic species, but recent progress has produced degenerate Fermi gases for a range of new species such as $^{173}$Yb~\cite{Takeshi2007}, $^{167}$Er~\cite{Aikawa2014},  $^{161}$Dy~\cite{Ravensbergen2020}, and $^{53}$Cr~\cite{Ciamei2022}. Thus, we anticipate an exciting time ahead for Fermi polarons, and for quantum mixtures in general. 

In the following, we present the theoretical description of Fermi polarons and how they relate to the highly imbalanced limit of two-component Fermi gases. Note that we do not attempt to provide an exhaustive survey of the literature, which can be found in recent reviews, e.g., Ref.~\cite{Scazza2022} and references therein. 
In Section~\ref{sec:model}, we introduce the model that describes a general two-component Fermi gas and briefly discuss the aspects of scattering theory and renormalisation that are of relevance to the Fermi polaron problem. In Section~\ref{sec:variational}, we describe a variational approach that can be used to calculate the Fermi polaron properties. Section~\ref{sec:ground} concerns the ground state behaviour of both a single impurity and many impurities, while Section~\ref{sec:dynamics} is focused on the various experimental probes of the static and dynamic properties of Fermi polarons. Finally, in Section~\ref{sec:outlook} we provide a brief outlook.

\section{Model of the two-component Fermi gas} \label{sec:model}

In the second quantised formalism~\cite{fetterbook}, the simplest model of the two-component Fermi gas in a volume $\vol$ is
\begin{align}
    \label{eq:Ham}
    \hat H=\underbrace{\sum_{\k,\sigma}\epsilon_{\k\sigma} \hat c^\dag_{\k\sigma} \hat c_{\k\sigma}}_{\hat H_0}+\underbrace{\frac g\vol\sum_{\k\k'\q}\hat c^\dag_{\k\up}\hat c_{\k+\q\up}\hat c^\dag_{\k'\down}\hat c_{\k'-\q\down}}_{\hat V}.
\end{align}
We use $\sigma=\up,\down$ to denote the two different types of fermions considered in these lectures, which can be either two hyperfine states of the same atom, or single hyperfine states of two different species of fermionic atoms. 
In the Hamiltonian~\eqref{eq:Ham}, the particles are described by the operators $\hat c^\dag_{\k\sigma}$ and $\hat c_{\k\sigma}$ that, respectively, create and annihilate a fermion of spin $\sigma$ and momentum $\hbar\k$. 
The two types of fermions have bare masses $m_\sigma$ and dispersions $\epsilon_{\k\sigma}=\hbar^2 k^2/2m_\sigma$, as well as particle number $N_\sigma$. The Fermi-polaron problem corresponds to the high-polarization limit $N_\down/N_\up \to 0$.  

The second term in Eq.~\eqref{eq:Ham} describes the underlying attractive  interparticle interactions. This can be precisely controlled in cold-atom experiments via the use of a magnetically tunable Feshbach resonance (see the review of Ref.~\cite{Chin2010} for further details), which has been crucial for the investigation of the strongly correlated Fermi gas~\cite{Zwerger2016}. While in realistic systems, the atoms interact via the van der Waals force, in practice the atoms collide at such small momenta that we are at liberty to replace the exact potential with a convenient attractive pseudo-potential that has the same low-energy behaviour. 
In particular, if we consider the case of a so-called broad Feshbach resonance, then the atom-atom scattering is parameterised by a single parameter---the $s$-wave scattering length $a$.\footnote{For the case of a so-called narrow Feshbach resonance, we also require an additional parameter---the effective range---which can be straightforwardly included in the theory by using a two-channel model~\cite{Timmermanns1999}.}
Thus, we can use the contact delta-function potential $g\delta(\mathbf{r})$ to describe the interaction between two distinguishable atoms at relative separation $\mathbf{r}$,  which in momentum space becomes simply the constant $g$ (which we set to zero above a relative momentum of $\hbar \Lambda$). This has the advantage that the two-particle scattering problem can be solved analytically, allowing us to easily relate the ``bare'' interaction strength $g$ to the physical scattering length $a$, as we show below.

\subsection{Scattering theory, $T$ matrix, and renormalization}\label{sec:scat}
Since the theoretical description of the quantum impurity problem described by the Hamiltonian~\eqref{eq:Ham} involves concepts of scattering theory and renormalization, we now briefly discuss the most relevant points of these theories, as they are often  
omitted in scientific papers. The reader who is already familiar with this formalism is, of course, welcome to skip this section.

\begin{figure}
\centering
\includegraphics[width=0.4\textwidth]{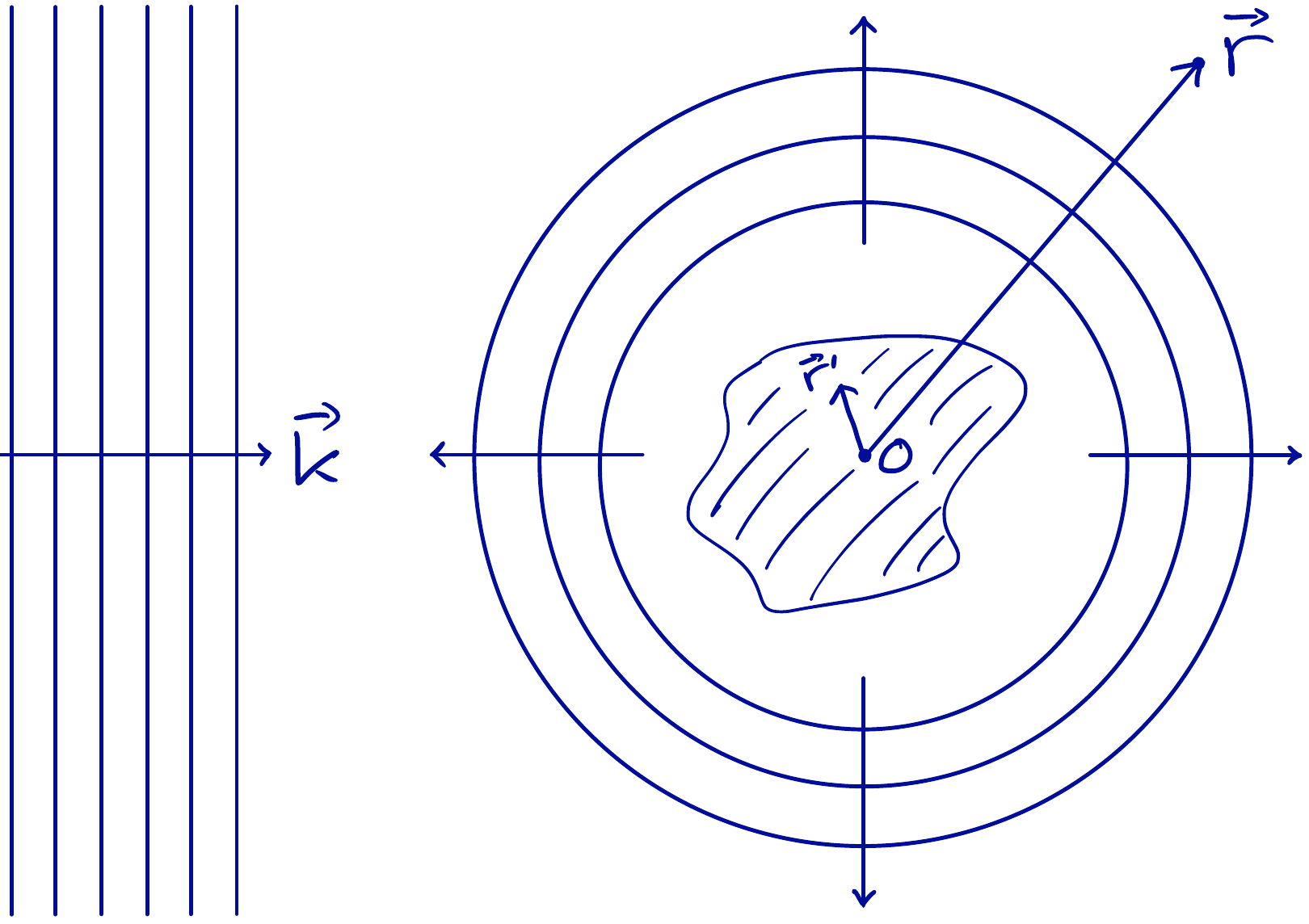}
\caption{Sketch of the steady-state scattering setup. We have a superposition of an incident plane wave with wave vector $\k$ and an expanding spherical wave, which both extend over all space. The scattering potential acts within the shaded region encompassing the origin. \label{fig:scatsketch}}
\end{figure}

We first aim to quantify the strength of a quantum scattering process for two particles ($N_\up = N_\down = 1$) via the so-called scattering amplitude. For simplicity, we work in the centre-of-mass frame since in a Galilean invariant system we can take the centre-of-mass momentum to vanish without loss of generality.
The problem then reduces to a one-body problem of a particle scattering off a potential at the origin. 
Furthermore, we consider elastic scattering of our spin $\up$ and $\down$ particles at momentum $\pm\hbar \k$ such that no energy is absorbed into other degrees of freedom during the collision, i.e., we have a well-defined energy $E=\epsilon_{\k\up}+\epsilon_{-\k\down}=\hbar^2k^2/2m_r$
with $m_r=1/(m_\up^{-1}+m_\down^{-1})$ the reduced mass.
While the scattering process is fundamentally time-dependent (see, e.g., Ref.~\cite{sakurai_napolitano_2020}), we will here formulate it as a steady-state problem where the total wave function consists of terms corresponding to an incident flux that is independent of the scattering, and an outgoing flux from the origin of the scattering potential, as sketched in Fig.~\ref{fig:scatsketch}. 

To this end, we take the incident wave function to be a plane wave $e^{i\k\cdot\rvec}$, where we do not worry about the overall normalization since only the ratio of incident to scattered wave amplitudes matter. In the presence of the scattering potential, but far outside the range of the potential\footnote{This asymptotic region can be defined for any potential that falls off faster than $1/r^2$~\cite{landau2013quantum}. The interaction potential we consider here certainly satisfies this requirement since the interatomic van der Waals potential scales as $1/r^6$ for large $r$.}, 
the outgoing wave function must satisfy two properties: (i) it has a constant probability current along any radial direction away from the origin such that there are no sources or sinks of probability, and (ii) it is an energy eigenstate. This implies that the total wave function has the asymptotic form 
\begin{align}
    \Psi(\rvec)=e^{i\k\cdot\rvec}+\frac{e^{ikr}}r f(\k',\k),
    \label{eq:scatsup}
\end{align}
for $\rvec$ far outside the range of the potential. We define $\k'$ as a wave vector of magnitude $k$ in the direction defined by $\rvec$. The quantity $f(\k',\k)$ is the scattering amplitude, which determines the strength of the scattering process since the ratio of the probability flux scattered into an infinitesimal solid angle $d\Omega$ along $\rvec$ relative to the incident probability flux is
\begin{align}
    \frac{d\sigma}{d\Omega}=|f(\k',\k)|^2.
\end{align}
This is called the differential cross section.

In order to calculate the scattering amplitude for a generic potential, we note that, formally, the solution to the \sch equation $(\hat H_0+\hat V)\kps=E\kps$ is the so-called Lippmann-Schwinger equation
\begin{align}
    \ket{\Psi}=\ket{\k}+\underbrace{\frac1{E-\hat H_0+i0}}_{\hat G_0(E)}\hat V\kps,
    \label{eq:LS}
\end{align}
where the incident-wave state $\ket{\k}\equiv \hat c^\dag_{\k\up}\hat c^\dag_{-\k\down}\ket{0}$, and
where $+i0$ denotes a small positive imaginary quantity $+i\epsilon$ in the limit $\epsilon\to0^+$.
This shift into the complex plane removes the divergence in Eq.~\eqref{eq:LS} due to $E$ coinciding with $\epsilon_{
\k\up}+\epsilon_{\k\down}$, an eigenvalue of $\hat{H}_0$, and it physically corresponds to an outgoing scattered wave ($-i0$ would give an incoming wave).

It can be easily checked that Eq.~\eqref{eq:LS} is a solution  by using the fact that 
the incident state satisfies  $\hat H_0\ket{\k}=E \ket{\k}$. To link Eq.~\eqref{eq:LS} back to the scattering amplitude, we consider its position-space representation\footnote{Here, we use the definition $\braket{\mathbf{r}}{\k} = e^{i\k\cdot\rvec}$, and the fact that the matrix element of the Green's operator is $\bra{\rvec'}\hat G_0(E)\ket{\rvec}=-\frac1{4\pi}\frac{2m_r}{\hbar^2}\frac{e^{ik|\rvec-\rvec'|}}{|\rvec-\rvec'|}\simeq -\frac1{4\pi}\frac{2m_r}{\hbar^2}\frac{e^{ikr}}{r}e^{-i\k'\cdot\rvec'}$~\cite{sakurai_napolitano_2020} where in the second step we perform a Taylor expansion assuming that the position $\rvec$ is far outside the range of the potential, while $\rvec'$ is close to the origin.}
\begin{align}
    \Psi(\rvec)&=e^{i\k\cdot\rvec}-\frac1{4\pi}\frac{2m_r}{\hbar^2}\int d^3r' \frac{e^{ik|\rvec-\rvec'|}}{|\rvec-\rvec'|}V(\rvec')\Psi(\rvec')\nn \\
    & \simeq e^{i\k\cdot\rvec}-\frac1{4\pi}\frac{2m_r}{\hbar^2}\frac{e^{ikr}}r\bra{\k'}\hat V\kps.
\end{align}
This already has the form in Eq.~\eqref{eq:scatsup}. Defining the transition operator $\hat T$ by $\hat T\ket{\k}=\hat V\ket{\Psi}$, we can formally identify the scattering amplitude as
\begin{align}
    f(\k',\k)=-\frac1{4\pi}\frac{2m_r}{\hbar^2}\bra{\k'}{\hat T}\ket{\k}.
    \label{eq:ffromT}
\end{align}
Furthermore, by acting on Eq.~\eqref{eq:LS} with $\hat V$ one finds that $\hat T$ satisfies the operator relation for given energy $E$,
\begin{align}
    \hat T (E)=\hat V+\hat V\hat G_0(E)\hat T(E).
    \label{eq:Top}
\end{align}
For certain interaction potentials, such as contact interactions, the matrix elements of the $T$ operator can be calculated analytically, which significantly simplifies the theory.

In ultracold atomic gases, the typical wavelength associated with the wave vector $\k$ (set by either the temperature or the interparticle spacing) is much larger than the characteristic size of the scattering potential, i.e., the van der Waals length. In this case, the scattering is insensitive to the precise shape of the potential and becomes spatially uniform, i.e., only $s$-wave scattering contributes. The scattering amplitude then takes the particularly simple form
\begin{align}
    f(\k',\k)=-\frac1{a^{-1}+ik},
    \label{eq:flowE}
\end{align}
that is, it is described by a single key parameter, the scattering length $a$. This is an excellent approximation for the scattering amplitude in the vicinity of the broad Feshbach resonances that are most commonly employed in cold-atom experiments~\cite{Chin2010}. 

The size and sign of the scattering length determine the effective low-energy interactions in quantum gases. In particular, close to a magnetically tunable Feshbach resonance, the scattering length behaves as~\cite{Chin2010}
\begin{align}
    a(B) = a_\mathrm{bg} \left(1-\frac\Delta{B-B_0}\right),
\end{align}
where $a_\mathrm{bg}$ is the background scattering length, $B_0$ is the magnetic field resonance position, and $\Delta$ is the resonance width. The special unitarity point $1/a = 0$, where the scattering length diverges, signifies the appearance of a two-body bound state (dimer) and corresponds to strong resonant interactions. For $a > 0$ there exists a $\up\down$ dimer with binding energy $\eb = \hbar^2/(2m_r a^2)$, which is linked to a pole of the scattering amplitude in Eq.~\eqref{eq:flowE}. In this case, the scattering states experience an effectively repulsive interaction due to the presence of the bound state, and the limit $a \to 0^+$ corresponds to an increasingly large binding energy $\eb \to \infty$ and vanishing interactions between scattering states. Conversely, for $a< 0$, there are only unbound scattering states with attractive $\up$-$\down$ interactions that vanish in the limit $a \to 0^-$. The key point to remember is that the underlying interaction potential is short ranged and attractive, no matter the value of $a$. 

\subsubsection{Contact interactions}
Thus far, our description of scattering theory applies to a generic short-range potential. We now consider the ultimate zero-range limit, namely the attractive delta-function potential in Eq.~\eqref{eq:Ham}, in which case the scattering never probes the detailed structure of the potential. Hence, the matrix element of the interaction potential reduces to a constant $\bra{\k_1}\hat V\ket{\k_2}=g$ for $k_1,k_2<\Lambda$, and 0 otherwise. We can then evaluate the scattering length by taking the limit of zero-energy scattering in Eq.~\eqref{eq:flowE} and using Eqs.~\eqref{eq:ffromT} and \eqref{eq:Top}:~\footnote{Here, in order to evaluate the matrix elements of the operator products in Eq.~\eqref{eq:Top} we insert the resolution of the identity, $\frac{1}{\vol}\sum_\k\ketbra{\k}$, after each $\hat G_0$ operator and use the facts that the matrix elements of $\hat V$ are constant up to $\Lambda$ while $\ket{\k}$ is an eigenstate of $\hat G_0$. Explicitly, for instance $\expval{\hat V\hat G_0(0)\hat V}{\0}=\frac{1}{\vol}\sum_\k\expval{\hat V\hat G_0(0)\!\ketbra{\k}\!\hat V}{\0}=-\frac{g^2}\vol\sum_\k^\Lambda \frac1{\epsilon_{\k\up}+\epsilon_{\k\down}}$ (we drop the $i0$ since the limit of zero imaginary part is well-defined). The sum is evaluated by taking the continuum limit $\frac1\vol\sum_\k\to \int \frac{d^3k}{(2\pi)^3}$. }
\begin{subequations}
\label{eq:afromg}
\begin{align}
\label{eq:afromga}
    a&= \frac1{4\pi}\frac{2m_r}{\hbar^2}\expval{\hat T}{\0}=\frac1{4\pi}\frac{2m_r}{\hbar^2}\left(g+g\Pi(0)g+g\Pi(0)g\Pi(0)g+\dots\right) \\
\label{eq:afromgb}
    &=\frac1{4\pi}\frac{2m_r}{\hbar^2}\frac1{g^{-1}-\Pi(0)}.
\end{align}
\end{subequations}
Here, we define $\Pi(0)\equiv -\frac1\vol\sum_\k^\Lambda\frac1{\epsilon_{\k\up}+\epsilon_{\k\down}}$, and the second line is obtained from the first by recognising the 
geometric series $1/(1-x)=1+x+x^2+\dots$.

At first sight, it is not obvious what we have gained by writing our key interaction parameter, the scattering length $a$, in terms of  $g$ and the function $\Pi(0)$. More worryingly, 
when we evaluate the sum appearing in $\Pi(0)$, we find that $\Pi(0)\propto -\Lambda$, which diverges as $\Lambda\to\infty$. 
How should we then understand Eq.~\eqref{eq:afromg}?

The resolution to this problem is to use the process of renormalization: The key idea is that $g$ and $\Lambda$ are \textit{bare} parameters of the model in the sense that they are not observables. Instead, the only physically meaningful parameter is the scattering length, which is experimentally measurable. Within the theory, for each value of $\Lambda$ we should therefore adjust the parameter $g$ in such a way that the scattering length has the desired value. 
In practice, one can use Eq.~\eqref{eq:afromg} to define $g$ in terms of $a$ and $\Lambda$ and, when this is done carefully, all divergent sums cancel and we are left with a theory that is well-defined when $\Lambda\to\infty$. Indeed, we shall see an explicit example below of how $g$ and $\Lambda$ can be completely removed from the final theory expressions. In this manner, the delta-function potential has precisely the same low-energy behaviour as the more realistic van der Waals interaction, while being much more theoretically tractable.

\section{Theoretical description of the polaron problem}
\label{sec:variational}

\begin{figure}
\centering
\includegraphics[width=.6\textwidth]{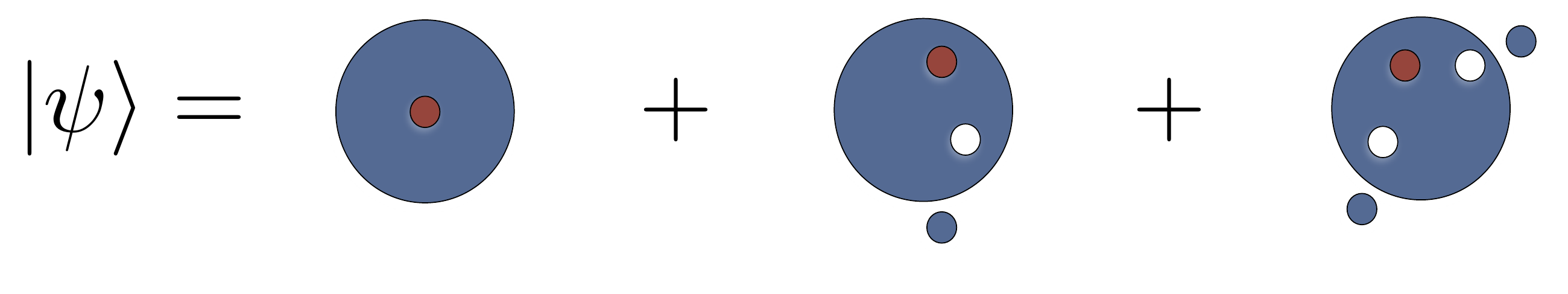}
\caption{Schematic illustration of the variational ansatz for the Fermi polaron consisting of the impurity (red circle) and the Fermi sea (large blue circle). From left to right we have 0, 1, or 2 particle-hole excitations (blue and white circles). \label{fig:sketch}}
\end{figure}

We now turn to the problem of a single $\down$ impurity immersed in a $\up$ Fermi sea with density $n_\up = N_\up/\vol$ and corresponding Fermi wave vector $k_F = (6\pi^2 n_\up)^{1/3}$ and Fermi energy $\ef = \hbar^2\kf^2/2m_\up$. Here, the scattering between the impurity and the surrounding fermions generates particle-hole excitations of the Fermi sea, as illustrated in Fig.~\ref{fig:sketch}. This in turn leads to the impurity becoming ``dressed'' by these medium excitations to form a Fermi polaron quasiparticle, a modified particle that still resembles the original bare impurity in the sense that it is adiabatically connected to the non-interacting state, but it has a different mass and other modified properties. There are two such polaron quasiparticles: the ``attractive'' Fermi polaron~\cite{Chevy2006}, which corresponds to the ground state at zero temperature, and the ``repulsive'' Fermi polaron~\cite{Cui2010,Massignan2011,Schmidt2011}, a metastable state at higher energy which connects to the scattering states as $a \to 0^+$. We will explore more deeply the concept of the polaron quasiparticle in the following sections.

Before doing so, we note that the single-impurity scenario is only a good description of the high-polarization limit, $N_\down/N_\up \to 0$, of a two-component Fermi gas provided that the minority $\down$ atoms can be treated as independent and uncorrelated. This can always be achieved at a non-zero temperature $T$ for a sufficiently small $\down$ density $n_\down = N_\down/\vol$, i.e., such that we have $\hbar^2 n_\down/m_\down \ll k_B T$, where $k_B$ is the Boltzmann constant. In this case, the $\down$ impurities form a classical Boltzmann gas even at low temperatures where the $\up$ Fermi gas is quantum degenerate, $T_F \equiv \ef/k_B \gtrsim T$. Thus far, most of the cold-atom experiments on Fermi polarons have effectively probed this Boltzmann regime where single-impurity theories are appropriate. However, for higher impurity densities or lower temperatures, one needs to consider the effective interactions between polarons when determining the applicability of the single-impurity limit. We will return to this point when we discuss the ground-state properties in Section~\ref{sec:ground}. 

\subsection{Variational approach to impurity dynamics}
The quantum impurity problem is a challenging many-body problem and thus we must resort to approximations. Since our aim is to describe impurities both in and out of equilibrium, a convenient method is to use a time-dependent variational principle~\cite{McLachlan64}, which provides a remarkably good  approximation of both the ground state and all excited states. Here, we will present the variational approach of Ref.~\cite{Parish2016} which is appropriate for a system at $T=0$; however we note that a similar variational method can also be formulated at finite temperature --- see Ref.~\cite{Liu2019}.

We assume that the system dynamics is governed by the time-independent Hamiltonian $\hat H$ and that we have a state $\ket{\psi(t)}$ that approximately describes how the system evolves as a function of time $t$. Of course, in the ideal case, $\ket{\psi(t)}$ would evolve exactly according to the \sch equation $i\hbar \partial_t\ket{\psi(t)}=\hat H\ket{\psi(t)}$. However, in practice we will incur errors due to not being able to describe all possible degrees of freedom. We can therefore define the ``error'' quantity~\cite{McLachlan64}
\begin{align} \label{eq:error}
 \Delta(t) =\expval{\hat \epsilon^\dag\hat \epsilon}{\psi(t)},
\end{align}
in terms of the error operator
$\hat{\epsilon} = i\hbar\partial_t - \hat H$. Our task is then to minimise this error quantity to obtain the most accurate system evolution.

Let us now assume that we evaluate the dynamics within a subset $\left\{\ket{j}\right\}$ of a complete and orthonormal set of states describing the impurity plus medium. We therefore have $\ket{\psi(t)} = \sum_j \alpha_j(t) \ket{j}$ where each $\alpha_j(t)$ is a time-dependent complex function. Now minimising the error incurred from time $t$ to $t+dt$ amounts to imposing the condition $\frac{\partial \Delta}{\partial\dot{\alpha}^*_j} = 0$ which implies 
\begin{align}
\bra{j} \hat{\epsilon} \ket{\psi(t)} = 0.
\end{align}
In other words, the optimal time evolution is obtained by projecting the \sch equation onto each basis state $\ket{j}$. Exploiting the orthonormality of the basis states,
$\bra{j}l \rangle = \delta_{jl}$, then finally yields the equations of
motion within the truncated subspace:
\begin{align}\label{eq:motion}
i\hbar  \frac{d\alpha_j}{dt} = \sum_l  \bra{j} \hat H \ket{l} \alpha_l(t).
\end{align}
This also allows us to determine the stationary states, $\alpha_j(t) = \alpha_j e^{-iEt/\hbar}$, which correspond to the energy eigenstates of the Hamiltonian in this subspace and which satisfy the corresponding equations
\begin{align} \label{eq:eigenstates}
   E \alpha_j = \sum_l  \bra{j} \hat H \ket{l} \alpha_l.
\end{align}
Note that the norm of the wave function is preserved also in the approximate evolution since the time evolution operator within the subspace is unitary.

\subsection{Variational ansatz and equations of motion} \label{sec:ansatz}

Let us now apply this approach to the Fermi polaron. The exact eigenstates of the system will involve an infinite series of all possible scattering processes between the impurity and the Fermi sea. 
Thus, a natural subset of these processes is 
to consider the lowest-order scattering terms, i.e., the term where the $\down$ impurity has left the $\up$ Fermi sea unperturbed, the terms where it has excited one particle-hole pair, and the terms with two particle-hole pairs, 
as illustrated in Fig.~\ref{fig:sketch}. The corresponding variational ansatz for an impurity energy eigenstate with zero total momentum 
is~\cite{Chevy2006,Combescot2008}
\begin{align} \label{eq:ansatz}
    \ket{\psi}=\Biggl(&\alpha_0 \, \hat{c}^\dag_{\0\down}+ \frac{1}{\vol}\sum_{\k\q}\alpha_{\k\q} \, 
    \hat c^\dag_{\q-\k\down}\hat c^\dag_{\k\up}\hat c_{\q\up} 
    \nn \\ &
    +\frac1{4\vol^2}\sum_{\k_1\k_2\q_1\q_2}\alpha_{\k_1\k_2\q_1\q_2} \, 
    \hat c^\dag_{\q_1+\q_2-\k_1-\k_2\down}\hat c^\dag_{\k_1\up}\hat c^\dag_{\k_2\up}\hat c_{\q_1\up}\hat c_{\q_2\up}
    \Biggl)\ket{\mathrm{FS}},
\end{align}
where $\ket{\mathrm{FS}}=\prod_{\q,q<k_F}\hat{c}^\dag_{\q\up}\ket{\rm vac}$ corresponds to the $\up$ Fermi sea ($\ket{\rm vac}$ is the vacuum).
Here, and in the following, we implicitly assume that all $\k$ and $\q$ wave vectors correspond, respectively, to particles and holes, i.e., $k>k_F$ and $q<k_F$.
Due to fermionic statistics, we require $\alpha_{\k_1\k_2\q_1\q_2}=-\alpha_{\k_2\k_1\q_1\q_2}=-\alpha_{\k_1\k_2\q_2\q_1}$. 

We now use Eq.~\eqref{eq:eigenstates} to derive the eigenstate equations for the variational ansatz. Importantly, since the interaction term $\hat V$ in the Hamiltonian \eqref{eq:Ham} describes the interaction of the impurity with a single majority particle, the number of particle-hole pairs can only be changed by at most one in the Hamiltonian matrix element in Eq.~\eqref{eq:eigenstates}. Thus we arrive at a hierarchy of linear equations~\cite{Chevy2006,Combescot2008}
\begin{subequations}
\label{eq:vareqs}
\begin{align}
\label{eq:vareqs-1}
    E_0\alpha_0&=\frac{g}{\vol^2}
    \sum_{\k\q}\alpha_{\k\q}
    ,\\ \label{eq:vareqs-2}
E_{\k\q}\alpha_{\k\q}&=g\Biggl(\alpha_0+
\frac{1}{\vol}\sum_{\k'}\alpha_{\k'\q}-
\frac{1}{\vol}\sum_{\q'}\alpha_{\k\q'}-
\frac{1}{\vol^2}\sum_{\k'\q'}\alpha_{\k'\k\q'\q}\Biggl),\\
E_{\k_1\k_2\q_1\q_2}\alpha_{\k_1\k_2\q_1\q_2}& = g\Biggl(\alpha_{\k_1\q_2}-\alpha_{\k_1\q_1}+\alpha_{\k_2\q_1}-\alpha_{\k_2\q_2}\nn \\ & \hspace{-22mm}+\frac{1}{\vol}\sum_{\k'}\left(\alpha_{\k'\k_2\q_1\q_2}
\alpha_{\k_1\k'\q_1\q_2} \right)
-\frac{1}{\vol}\sum_{\q'}\left(\alpha_{\k_1\k_2\q'\q_2}
+\alpha_{\k_1\k_2\q_1\q'} \right)
+\dots\Biggl),
\end{align}
\end{subequations}
where $\dots$ indicates terms arising from states with \emph{three} particle-hole excitations, which do not appear in our variational ansatz and can thus be omitted. In these equations,
$E_0=E-gn_\up$, $E_{\k\q}=E-gn_\up-\epsilon_{\k\up}+\epsilon_{\q\up}-\epsilon_{\q-\k\down}$, and $E_{\k_1\k_2\q_1\q_2}=E-gn_\up-\epsilon_{\k_1\up}-\epsilon_{\k_2\up}+\epsilon_{\q_1\up}+\epsilon_{\q_2\up}-\epsilon_{\q_1+\q_2-\k_1-\k_2\down}$ denote the energy measured from the Hartree shift $gn_\up$ and the kinetic energies of the particles in the corresponding term of the variational ansatz, while the right hand side corresponds to non-trivial interaction effects due to impurity-medium scattering. Clearly, the equations rapidly increase in complexity when including multiple particle-hole excitations, and indeed most calculations have focused on the case of at most a single excitation.

In the continuum limit, the variational equations in~\eqref{eq:vareqs} constitute a set of coupled inhomogeneous Fredholm integral equations of the second kind that can be solved by applying a quadrature rule to approximate the integrals on a grid, effectively converting it into a matrix equation~\cite{NumericRec}. Note that some care must be taken in normalising the equations according to $\abs{\alpha_0}^2+\frac1{\vol^2}\sum_{\k\q}\abs{\alpha_{\k\q}}^2+\frac1{4\vol^4}\sum_{\k_1\k_2\q_1\q_2}\abs{\alpha_{\k_1\k_2\q_1\q_2}}^2=1$. In addition, the equations still suffer from the problem of ultraviolet divergences and hence, for a given value of the dimensionless interaction strength $1/k_Fa$, one must choose a cutoff $\Lambda\gg k_F,\abs*{a^{-1}}$ and relate $g$ to $\Lambda$ using Eq.~\eqref{eq:afromg}. In this manner, one obtains well-defined results in the limit $\Lambda\to\infty$.

To see that the limit of $\Lambda\to\infty$ is indeed well defined, we can systematically eliminate the cutoff dependence and replace $g$ by the physically measurable parameter $a$~\cite{Chevy2006}. To see how this works, we will consider only the case of a single particle-hole excitation, but the idea is similar for multiple excitations~\cite{Combescot2008}. Define
\begin{align}
    \chi_\q&=\frac g{\alpha_0\vol}\sum_\k \alpha_{\k\q}.
\end{align}
Since $\alpha_{\k\q}$ scales as $1/k^2$ at large $k$ [see Eq.~\eqref{eq:vareqs-2}], $\chi$ is finite in the limit $\Lambda\to\infty$. By manipulating Eqs.~\eqref{eq:vareqs-1} and \eqref{eq:vareqs-2} we then find 
\begin{subequations}
\label{eq:Eequalsstuff}
\begin{align} 
    E&=gn_\up+\frac1\vol\sum_\q\chi_\q \\
    & =\cancelto{0}{gn_\up}\, \,\, \,\, \, + \,\frac1\vol\sum_\q\left[\frac1g-\frac1\vol\sum_\k\frac1{E_{\k\q}}\right]^{-1}\cancelto{1}{\left[\frac g\vol\sum_\k 
    \frac1{E_{\k\q}}\left(1-\frac 1{\vol}\sum_{\q'}\frac{\alpha_{\k\q'}}{\alpha_0}\right)\right]}\\
    &\!\!\! \stackrel{\Lambda\to\infty}{\longrightarrow} \frac1\vol\sum_\q\left[\frac{m_r}{2\pi\hbar^2a}\!-\!\frac{m_r k_F}{\pi \hbar^2}\!-\!\frac1\vol\sum_\k\left(\frac1{\epsilon_{\k\up}+\epsilon_{\k\down}}\!-\!\frac1{\epsilon_{\k\up}-\epsilon_{\q\up}+\epsilon_{\q-\k\down}-E}\right)\right]^{-1} ,\label{eq:Erenorm}
\end{align}
\end{subequations}
where in the second line the arrows indicate the behaviour of the corresponding terms in the limit $\Lambda\to\infty$, and in the third line we have set the Hartree shift in $E_{\k\q}$ to zero and used Eq.~\eqref{eq:afromg} to replace $g$ with the scattering length. Now we see that the sum in Eq.~\eqref{eq:Erenorm} is convergent, and 
thus all cutoff dependence is removed from the problem.

Equation~\eqref{eq:Eequalsstuff} precisely corresponds to that derived within the so-called ladder (or non-self-consistent $T$-matrix) approximation~\cite{Combescot2007} for the ground-state energy. This should not come as a surprise, since our variational ansatz precisely consisted of those terms where the impurity repeatedly interacts with a single particle excited out of the Fermi sea, corresponding to the ``ladder'' terms in a diagrammatic formulation. Importantly, the variational approach may be used to effectively sum  classes of diagrams of increasing complexity in a systematic manner, for instance by going beyond the ladder diagrams and introducing two particle-hole excitations as in Eq.~\eqref{eq:ansatz}. 

While the solution to Eq.~\eqref{eq:Eequalsstuff} yields the ground-state energy, the coupled variational equations \eqref{eq:vareqs} yield the full spectrum of energies and associated eigenstates. Therefore, an obvious question is what happened to all the excited states? The derivation of Eq.~\eqref{eq:Eequalsstuff} implicitly assumes that $\alpha_0$ remains finite as $\vol \to \infty$ and there are no singularities in the integrals. However, this is not the case above the ground state, where there is a continuum of excited states, each with vanishing $\alpha_0$. Instead, one must treat it like the scattering problem in Section~\ref{sec:scat}, which amounts to considering the right hand side of Eq.~\eqref{eq:Eequalsstuff} as the impurity self energy and taking $E\to E+i0$~\cite{Scazza2022}. The details are beyond the scope of these lectures; suffice to say that it is completely equivalent to the ladder approximation. Likewise, when extending the variational approach based on a single particle-hole excitation to finite temperature~\cite{Liu2019}, one finds that the resulting set of equations can be manipulated to obtain the same self energy derived using finite-temperature ladder diagrams in the limit of uncorrelated impurities~\cite{Hu-Mulkerin_PRA2018,Tajima2018,Tajima2021}.

\begin{figure}
\centering
\includegraphics[width=\textwidth]{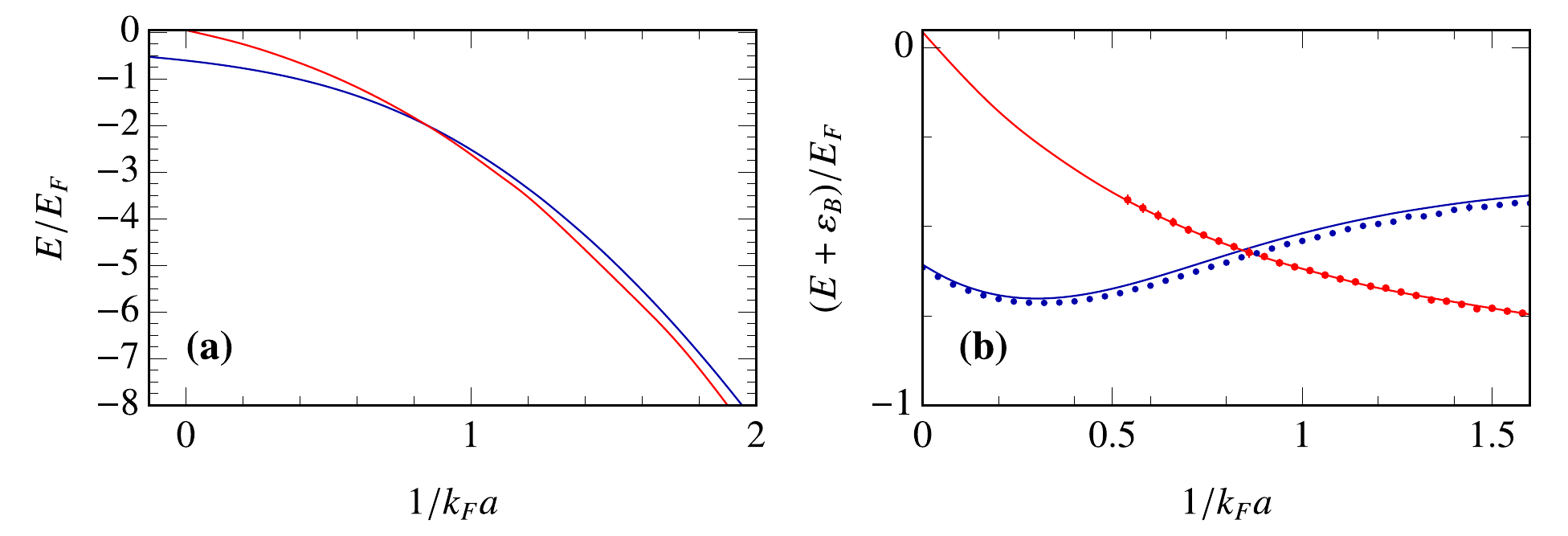}
\caption{Energies of the polaron (blue) and molecule (red) for equal masses $m_\down = m_\up$. These are calculated within the variational ansatz with one particle-hole excitation (solid lines)~\cite{Chevy2006,Combescot2007,Punk2009,Combescot2009} or using diagrammatic QMC (symbols, most error bars are smaller than the symbol size)~\cite{Vlietinck2013}. Panel (a) shows only the results of the variational ansatz, while panel (b) is a comparison between the two methods in the region around the polaron-molecule transition at $1/\kf a \simeq 0.88$~\cite{Combescot2009} where the two-body binding energy has been subtracted for clarity. \label{fig:polaron_results}}
\end{figure}

\section{Ground-state behaviour} \label{sec:ground}

Considerable insight can be gained from studying the polaron ground state, the lowest energy eigenstate of the many-body system. In the absence of interactions, the ground state is known exactly and simply corresponds to $\ket{\psi_0} = \hat{c}^\dag_{\0\down} \ket{\mathrm{FS}}$, which has zero energy with respect to the energy of the unperturbed Fermi sea. We can then imagine gradually perturbing away from this bare non-interacting state as we turn on impurity-fermion interactions, and we can quantify how much the interacting ground state $\ket{\psi}$ resembles the original unperturbed state by taking the squared overlap $Z = |\! \braket{\psi_0}{\psi} \!|^2 = |\alpha_0|^2$. Such a quantity is also known as the \emph{quasiparticle residue} and is one of several quasiparticle properties that characterise the Fermi polaron, along with the polaron energy $E$~\cite{Massignan2014,Scazza2022}. 

\begin{figure}
\centering
\includegraphics[width=\textwidth]{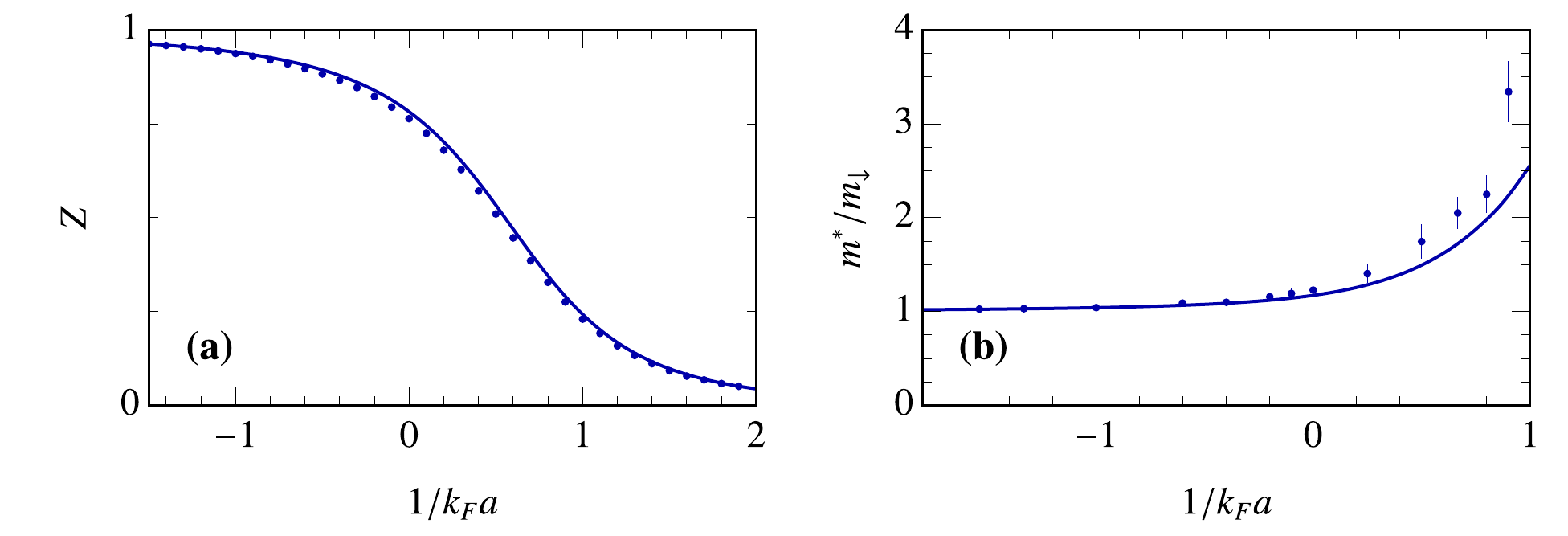}
\caption{(a) Polaron residue and (b) polaron effective mass, calculated within the polaron variational ansatz with a single-particle hole excitation (solid line) and within diagrammatic QMC (symbols, most error bars are smaller than the symbol size)~\cite{Vlietinck2013}. The masses are once again taken to be equal, $m_\down = m_\up$.  \label{fig:polaron_results2}}
\end{figure}

In the limit of weak attraction $\kf a \to 0^-$, the polaron still closely resembles the bare impurity such that its residue $Z$ is close to 1 (assuming the impurity mass $m_\down$ is finite). Furthermore, its energy is lowered and corresponds to a simple mean-field energy shift $E \simeq 2\pi \hbar^2 a n_\up/m_r$ at lowest order in $\kf a$. 
In this regime, we expect an ansatz like Eq.~\eqref{eq:ansatz}, with only a small number of particle-hole excitations of the Fermi sea, to provide a good approximation for the ground state. 
In particular, the simplest ansatz involving a single particle-hole excitation --- the so-called ``Chevy ansatz''~\cite{Chevy2006} --- correctly reproduces the energy and other quasiparticle properties obtained from perturbation theory up to order $(\kf a)^2$~\cite{Bishop1973,Trefzger2013}.
However, remarkably, the Chevy ansatz has been found to remain accurate all the way up to resonant interactions at unitarity and beyond: as shown in Figs.~\ref{fig:polaron_results} and \ref{fig:polaron_results2}(a), both the ground-state energy and quasiparticle residue are in excellent agreement with state-of-the-art diagrammatric quantum Monte Carlo (QMC) calculations~\cite{Vlietinck2013} for the case of equal masses $m_\down = m_\up$. 

Why is the simple variational approach so accurate? Firstly, it has been argued that the Fermi statistics of the medium leads to a near cancellation of higher order terms involving more than one particle-hole excitation~\cite{Combescot2008}. To see this, consider Eq.~\eqref{eq:vareqs-2} in the hierarchy of coupled linear equations for the polaron wave function and then perform a sum over the hole wave vector $\q$. This causes the term involving two particle-hole excitations to vanish due to the antisymmetry of the coefficient $\alpha_{\k_1\k_2\q_1\q_2}$ with respect to hole exchange. Now in principle this manipulation does not gain us anything 
since we are now left with a new unknown function $\sum_\q E_{\k\q} \alpha_{\k\q}$. However, if we assume that the coefficient $\alpha_{\k\q}$ is rather insensitive to $\q$, which is reasonable because of the phase space restriction $q < \kf$, then we can factor it out of the sum and we are back to a set of coupled equations we can solve.
Thus, it is a combination of the small hole momentum and the antisymmetry arising from Fermi statistics that almost completely decouples the subspace involving a single particle-hole excitation from all the other subspaces with multiple excitations.

Another feature of the Chevy ansatz is that it contains the two-body bound state, which has the form
$\ket{\psi_{2b}} = \frac{1}{\sqrt{\vol}}\sum_\k \varphi_\k \hat{c}^\dag_{-\k\down} \hat{c}^\dag_{\k\up} \ket{0}$ and which should dominate in the opposite limit of strong attraction $1/\kf a \to \infty$ when the impurity is sufficiently heavy that there are no other bound clusters.  
Thus, the ansatz also captures the leading order behaviour of the energy in this regime, given by $E \simeq - \eb \to -\infty$. This is in some sense a trivial limit that is independent of the medium density.  
However, studies of trapped few-body systems~\cite{Blume2008,Levinsen2017} have shown that, even at unitarity, the polaron energy for the many-body system is well approximated by just a few particles, implying that two-body physics (i.e., two-point correlations) dominates for all $1/\kf a$. This further justifies the use of variational approaches based on a small number of particle-hole excitations.

\subsection{Ground-state transitions}
While the Chevy ansatz correctly captures the binding energy of the bound $\up\down$ molecule, it fails to describe the many-body corrections involving the medium density in the limit $1/\kf a \to \infty$. This includes even the simple leading order correction, $-\ef$, which arises from the fact that a fermion must be removed from the $\up$ Fermi sea to form the dimer. 
The origin of this discrepancy is that the Chevy ansatz effectively assumes that there is always a well-defined polaron quasiparticle with $Z \neq 0$, as depicted in Fig.~\ref{fig:polaron_results2}(a). However, as was first shown using diagrammatic QMC~\cite{Prokofev2008} for equal masses, the impurity eventually undergoes a sharp transition 
to a molecule (dimer) state with increasing $1/\kf a$ (Fig.~\ref{fig:polaron_results}). 
Such a molecule has vanishing $Z$ and is well described by the modified variational ansatz~\cite{Punk2009,Mora2009}
\begin{align} \label{eq:mol-ansatz}
    \ket{\psi'} = \left( \frac{1}{\sqrt{\vol}} \sum_\k \varphi_\k  \, \hat{c}^\dag_{-\k\down} \hat{c}^\dag_{\k\up} + \frac{1}{2\vol^{3/2}}\sum_{\k_1\k_2 \q} \varphi_{\k_1\k_2\q} \, \hat c^\dag_{\q-\k_1 - \k_2\down}\hat c^\dag_{\k_1\up}\hat c^\dag_{\k_2\up}\hat c_{\q\up}\right)\ket{\mathrm{FS}'}.
\end{align}
Here, the state $\ket{\mathrm{FS}'}$ corresponds to the same $\up$ Fermi sea as before but with one less fermion, and the dressing cloud of the molecule is again approximated by a single particle-hole excitation. This yields the lowest order many-body corrections in the limit $1/\kf a \to \infty$, corresponding to the energy $E \simeq -\eb -\ef + 2\pi \hbar^2 a_{ad}n_\up/2m_{ad}$~\cite{Combescot2009}, where $a_{ad}$ is the scattering length for the interaction between the dimer and a $\up$ fermion~\cite{Petrov2003}, and $m_{ad}= m_\up (m_\up+m_\down)/(m_\down +2m_\up)$. 

We see that the dressed molecule $\ket{\psi'}$ is qualitatively different from the polaron quasiparticle, i.e., it is a bosonic rather than fermionic quasiparticle and it has zero overlap with the bare non-interacting state $\ket{\psi_0}$. Moreover, no parts of this state are contained in the Chevy ansatz.\footnote{Note that if we instead allow the Chevy ansatz to have a finite rather than zero momentum, then the ansatz with momentum $\hbar \kf$ contains the first bare dimer term in Eq.~\eqref{eq:mol-ansatz}~\cite{Parish2021}.} In order to convert a zero-momentum polaron into a zero-momentum molecule, one needs at least two additional particle-hole excitations 
to conserve momentum~\cite{Bruun2010}, leading to an abrupt transition and the associated kink in the energy in Fig.~\ref{fig:polaron_results}. 
Indeed, the ansatz \eqref{eq:ansatz} with up to two particle-hole excitations can describe this transition but only to the bare undressed molecule state [first term in Eq.~\eqref{eq:mol-ansatz}].  
We emphasise that such a sharp ``first-order'' transition is a feature of the many-body system in the thermodynamic limit $\vol \to \infty$, $N_\up \to \infty$, and does not appear in the few-body limit where it is instead replaced by a crossover. 
Furthermore, it is easily destroyed by thermal fluctuations~\cite{Ness2020,Parish2021}, but signatures remain at finite temperature (see Section~\ref{sec:dynamics}).

In addition to the residue $Z$, another property that characterises the transition and the nature of the impurity is the \textit{effective mass} $m^*$, which defines the response of the ground state to a change in impurity momentum $\p$, i.e., $1/m^* \equiv 
\left.\partial^2_{\p}E\right|_{\p=\0}$. In the absence of interactions, we simply have $m^* = m_\down$, while increasing $1/\kf a$ dresses the impurity and leads to a larger $m^*$, as shown in Fig.~\ref{fig:polaron_results}(b). This is again well described by the Chevy ansatz, with significant deviations from diagrammatic QMC only appearing as we approach the polaron-molecule transition. Near perfect agreement can be obtained, however, if one uses the ansatz \eqref{eq:ansatz} with two particle-hole excitations~\cite{Combescot2009,Vlietinck2013}. 

Close to the transition, we see in Fig.~\ref{fig:polaron_results} that the effective mass for the polaron quasiparticle can even become larger than the mass of a bare dimer (corresponding to $2 m_\down$ for the case of equal masses). If one continues along the polaron branch, one eventually finds that $m^*$ diverges and changes sign~\cite{Combescot2009}, signifying an instability of the ground state. In reality, before we reach this point, there is a transition to a dressed molecule with an abrupt increase in $m^*$, after which it monotonically decreases towards the bare dimer mass with increasing $1/\kf a$.

The ground state becomes even richer for a light impurity, where there is the prospect of other bound clusters involving the impurity such as three-body bound states (trimers), depending on the mass ratio $m_\up/m_\down$. In particular, there exists a universal trimer that is independent of the short-distance physics 
when $m_\up/m_\down \gtrsim 8.2$~\cite{Kartavtsev_2007}. 
The existence of such bound states leads to further transitions where the impurity quasiparticle changes its character, and the presence of the Fermi sea can even favour the formation of dressed trimer quasiparticles at lower mass ratios~\cite{Mathy2011}. This interesting scenario should soon be accessible experimentally due to the variety of atomic mixtures that are becoming available.

\subsubsection{Many impurities} Our discussion thus far has focused on the ground-state properties in the single-impurity limit. Here the statistics of the impurity is unimportant and thus it could correspond to the limit of a Bose-Fermi mixture as well as a Fermi-Fermi mixture. 
However, the nature of the impurity quasiparticle has implications for the many-body phases at finite impurity density. For the two-component Fermi gas considered in these lecture notes, 
the bosonic dressed molecules will form a Bose-Einstein condensate (BEC) while the fermionic polarons will form a Fermi liquid.\footnote{In principle, polarons can pair up to form a $p$-wave fermionic superfluid, but the temperature scale for this phase is essentially inaccessible in the limit of low impurity density, so we ignore this possibility in the following.} Furthermore, while there are no interactions between bare $\down$ impurities, there are medium-induced interactions between the impurity quasiparticles, which affect the stability of the single-impurity limit. Specifically, one can show that the dressed molecules are unstable towards collapse into a higher density gas~\cite{Chevy2010,Parish2021}, thus causing the single-impurity picture to fail at the polaron-molecule transition. One can understand this by considering the effective free energy of dressed (bosonic) molecules at low density $n_\down$: $F = E_\down n_\down + \frac{1}{2}U n_\down^2$, where $E_\down$ is the energy of a single impurity and $U$ is the interaction between impurities. In the limit, $1/\kf a \to \infty$, $U$ is simply given by the dimer-dimer interaction, which is repulsive in the equal-mass case~\cite{Petrov2004}. However, as we approach unitarity, $U$ eventually changes sign due to the attractive interactions induced by the $\up$ Fermi gas and thus the molecular BEC becomes unstable. Physically, this corresponds to phase separation between the BEC and the $\up$ Fermi gas, which dominates the phase diagram of the spin-imbalanced Fermi gas and which is known to preempt the polaron-molecule transition at high polarisation~\cite{Pilati2008,Radzihovsky2010}. 

Finally, we note that there is also the possibility of forming bound states involving a $\up$ fermion and two or more impurities when the mass ratio $m_\up/m_\down$ is sufficiently small~\cite{Kartavtsev_2007}, once again highlighting the additional rich physics in mass-imbalanced mixtures. 

\begin{table}
  \begin{tabular}{lll}
  & Fermi polaron  & Bose polaron \\
    \hline
    Medium & Non-interacting Fermi gas & Weakly repulsive BEC \\
      Dressing cloud      & Particle-hole excitations & Bogoliubov excitations \\
      Ground state & Abrupt transitions ($m_\down$ finite)  & Smooth crossover \\
      & Orthogonality catastrophe ($m_\down\to\infty$) & Analogue of OC ($a_B\to0)$   \\
    Unitarity limit & Scale invariance & Efimov bound states \\
    \hline
  \end{tabular}
  \caption{Comparison between Fermi and Bose polarons.}
  \label{tab:polarons}
\end{table}

\subsection{The role of the medium} 

We conclude this section with a discussion of the role of the medium and how the physics changes if we instead have a \emph{Bose polaron} --- an impurity immersed in a BEC~\cite{Scazza2022}. This scenario has been realised in several recent experiments, for instance Refs.~\cite{Hu2016,Jorgensen2016}. In this case, the impurity is dressed by Bogoliubov excitations of a BEC~\cite{Tempere2009,Rath2013,Li2014} rather than particle-hole excitations of a Fermi gas (see Table~\ref{tab:polarons}). While both types of excitations correspond to density fluctuations of the background medium, the underlying bosonic statistics in the BEC leads to qualitatively different behaviour near unitarity $1/a \to 0$. In particular, the Bose polaron does not exhibit abrupt transitions like the Fermi-polaron case, since an impurity in a BEC cannot change its statistics and thus its character by binding a boson. Furthermore, the Bose polaron features additional correlations arising from Efimov bound states~\cite{Naidon2017} involving the impurity plus two or more bosons, which means that the ground-state properties depend on an Efimov length scale (the three-body parameter) as well as the scattering length $a$~\cite{Levinsen2015b,Yoshida2018}. Thus, the Bose polaron is not scale invariant when $1/a = 0$, and still depends on an interaction length scale, which can in principle be set by the boson-boson scattering length $a_B$~\cite{Ardila2016,Yoshida2018}. Finally, while the classic orthogonality catastrophe (OC) for $m_\down \to \infty$ requires a Fermi medium~\cite{Anderson1967}, there is an analogue of OC when $a_B \to 0$, where $Z \to 0$ for any impurity-medium interaction due to the infinite compressibility of the ideal Bose gas~\cite{Shchadilova2016,Yoshida2018,Mistakidis2019,Guenther2021}. 

\section{Energy spectrum and dynamics}
\label{sec:dynamics}

\begin{figure}
\centering
\includegraphics[width=.6\textwidth]{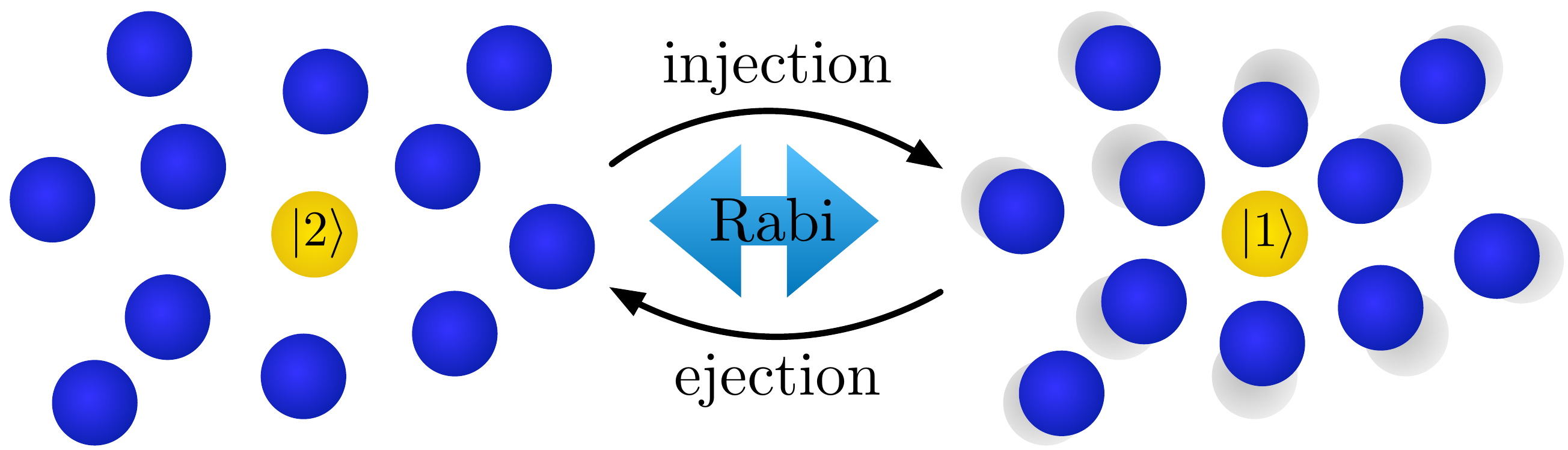}
\caption{Illustration of three different experimental protocols to probe the Fermi polaron, which all require an auxillary non-interacting state $\ket{2}$ of the impurity in addition to the state $\ket{1}$ with tunable medium-impurity interactions.
Rabi oscillations are achieved by coupling the two impurity states with a continuous drive (e.g., a radiofrequency field), while injection and  ejection spectroscopy probe the linear response of the impurity to a weak drive. 
Adapted from~\cite{Liu2020pra}. \label{fig:protocols}}
\end{figure}

We now turn to the physics beyond the ground state, namely the full spectrum of excited states and the dynamics of impurities driven out of equilibrium. Here, our aim is to give a brief overview of the protocols that can be realised in cold-atom experiments and our theoretical understanding of them based on the variational approach outlined in Section~\ref{sec:variational}. As illustrated in Fig.~\ref{fig:protocols}, probing the energy spectrum and dynamics requires an additional internal state of the impurity (e.g., another hyperfine state), which is non-interacting and acts as a reference for the interacting impurity system, allowing one to access polaron properties such as the residue, 
as well as excited states in the spectrum such as the repulsive polaron. 

\subsection{Injection and ejection spectroscopy}
If we start with the non-interacting impurity state $\ket{2}$ and then weakly drive it into the interacting state $\ket{1}$ at energy $\hbar \omega$ with respect to the bare transition, we can obtain the transfer rate using Fermi's golden rule:
\begin{align} \label{eq:injection}
    \mathcal{I}_{\rm inj}(\omega) = 
    \hbar \sum_j |\!\braket{\psi_0}{\gamma_j}\!|^2 \delta(\hbar \omega - E_j) \, .
\end{align}
Here $\ket{\gamma_j}$ and $E_j$ respectively correspond to the eigenstates and eigenvalues of the interacting system, and we have dropped the constant prefactors so that $\int d\omega \, \mathcal{I}_{\rm inj}(\omega) = 1$. For the variational ansatz in Section~\ref{sec:ansatz}, we simply have $|\!\braket{\psi_0}{\gamma_j}\!|^2 = |\alpha_0^{(j)}|^2$, where $j$ labels the approximate eigenstates and energies, i.e., the solutions to Eq.~\eqref{eq:vareqs}. Thus, the so-called ``injection'' spectroscopy  (Fig.~\ref{fig:protocols}) provides a measure of the energy spectrum weighted by the overlap with the non-interacting state. Equation~\eqref{eq:injection} corresponds to the definition of the impurity \textit{spectral function}~\cite{fetterbook,Punk2007}, which can also be obtained from diagrammatic approaches---e.g., the ladder approximation, which is equivalent to the lowest-order (Chevy) variational approach in the continuum limit, $\vol \to \infty$ (see Section~\ref{sec:ansatz}). 

\begin{figure}
\centering
\includegraphics[width=.7\textwidth]{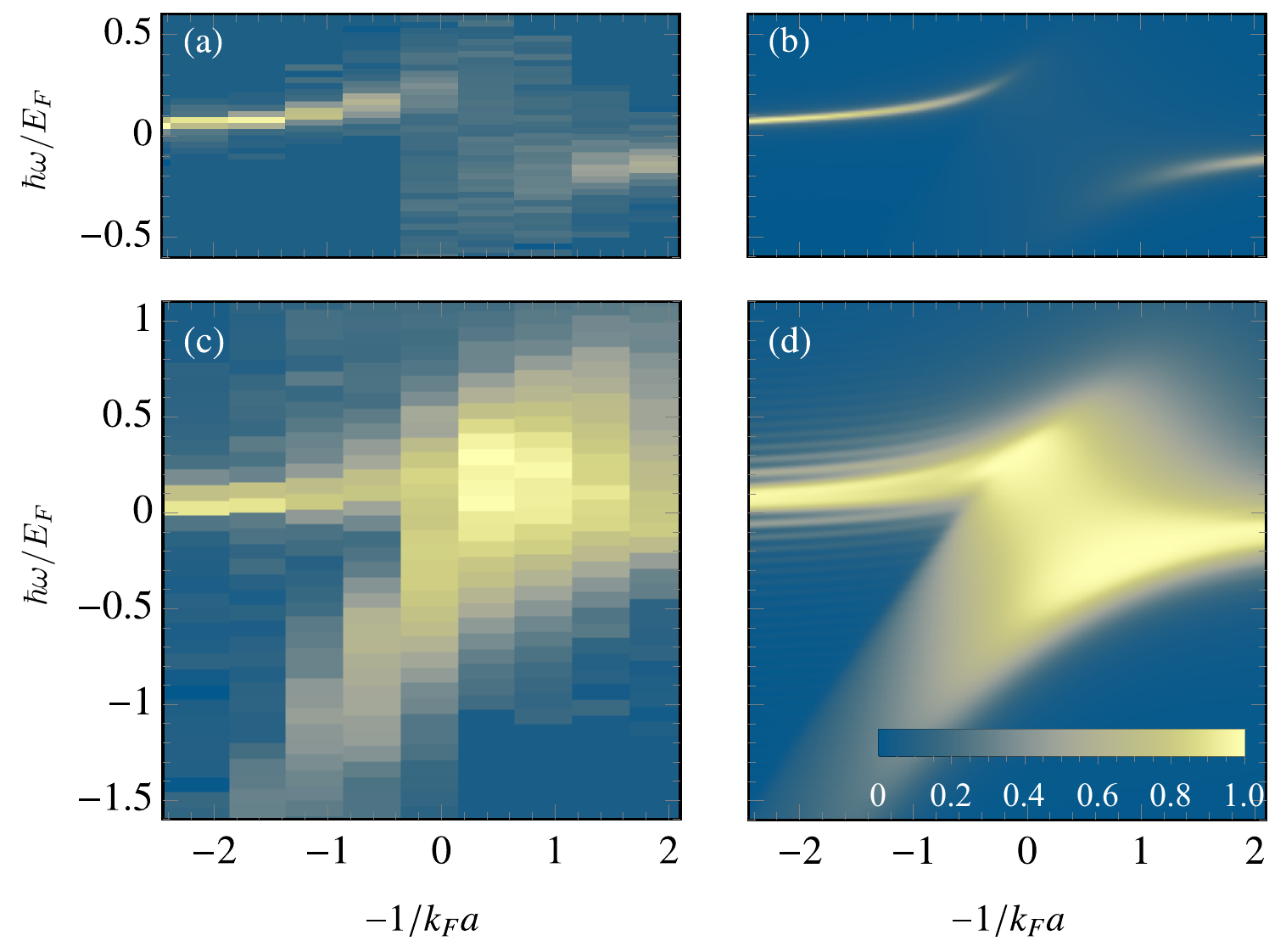}
\caption{Injection spectrum of $^{40}$K impurities in a $^6$Li Fermi sea at $T\simeq 0.16T_F$ as a function of interaction strength, obtained by applying a radiofrequency pulse. (a) and (c) show the experimental results of Ref.~\cite{Kohstall2012} in the linear response regime and for a strongly Rabi driven impurity, respectively. Panels (b) and (d) show the corresponding spectral response calculated by simulating the full time dynamics of a Rabi-driven infinitely heavy impurity~\cite{Adlong2021}. Adapted from~\cite{Adlong2021}. \label{fig:rf}}
\end{figure}

Figure~\ref{fig:rf} shows a typical injection spectrum (i.e., the transfer rate to the interacting state) for mobile impurities (e.g., $^{40}$K) in a degenerate atomic Fermi gas such as $^6$Li. In the linear response regime of a weak drive, where Eq.~\eqref{eq:injection} should be valid, one finds two dominant branches: the attractive polaron at negative energies (connected to the ground state discussed in Section~\ref{sec:ground}) and the metastable repulsive polaron at positive energies. These are adiabatically connected to the bare non-interacting impurity in the limits $1/\kf a \to -\infty$ and $1/\kf a \to \infty$, respectively. There is also a continuum of many-body states between these polaronic branches, which are dominated by dressed dimer states near unitarity, but the overlap with the non-interacting impurity is small and thus it is difficult to see within linear response [Figs.~\ref{fig:rf}(a,b)]. However, it becomes much more visible when the strength of the drive is large [Figs.~\ref{fig:rf}(c,d)], since the response is no longer simply proportional to the overlaps $|\alpha_0^{(j)}|^2$ and the drive itself also slightly changes the energy spectrum. The experimental observations are well reproduced by theoretical simulations based on a modified variational approach for an infinitely heavy impurity~\cite{Adlong2021}. 

For simplicity, we have focused on the zero-temperature limit in Eq.~\eqref{eq:injection}, but this can be straightforwardly generalised to non-zero $T$ by performing a thermal average over the initial states of the medium and the impurity~\cite{Liu2020prl,Liu2020pra}. Crucially, this allows one to relate $\mathcal{I}_{\rm inj}(\omega)$ to the transfer rate $\mathcal{I}_{\rm ej}(\omega)$ for \emph{ejection spectroscopy}, where an interacting impurity is weakly driven into a non-interacting state (Fig.~\ref{fig:protocols}), giving~\cite{Liu2020prl,Liu2020pra}
\begin{align} \label{eq:ejection}
    \mathcal{I}_{\rm ej}(\omega) = e^{\beta\hbar\omega} e^{\beta \Delta\mathcal{F}} \mathcal{I}_{\rm inj}(-\omega) \, 
\end{align}
where $\beta \equiv 1/k_B T$, and $\Delta\mathcal{F}$ is the difference in the impurity free energies for the interacting and non-interacting systems. Here we also have $\int d\omega \, \mathcal{I}_{\rm ej}(\omega) = 1$. Note that this relation only holds for an initial state in thermal equilibrium and for uncorrelated impurities, where $\beta \hbar^2 n_\down/m_\down \ll 1$.

\begin{figure}
\centering
\includegraphics[width=.5\textwidth]{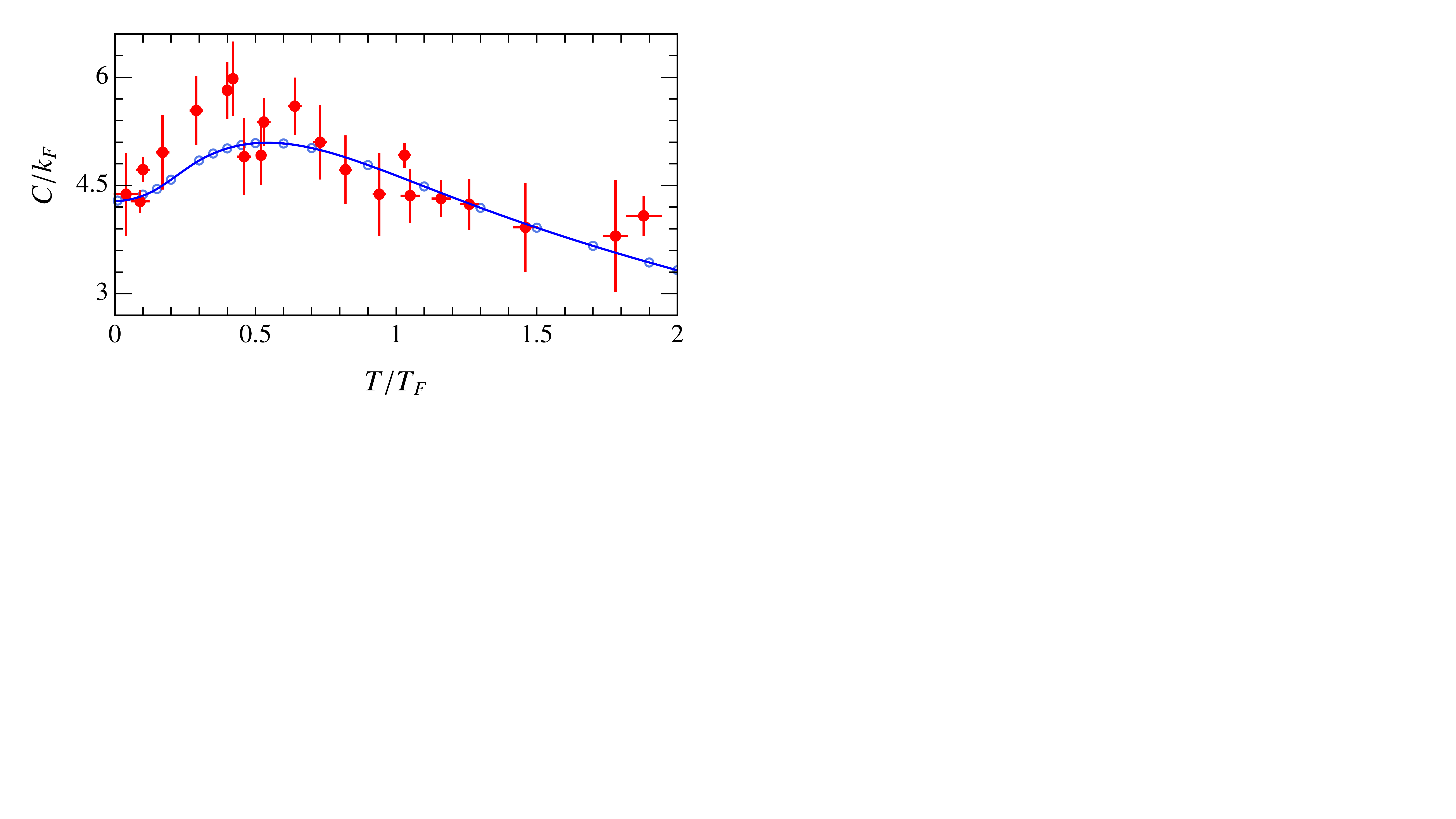}
\caption{Contact at unitarity as a function of temperature. Comparison between theory~\cite{Liu2020prl} and experiment~\cite{Yan2019} for a two-component $^6$Li gas. Adapted from~\cite{Liu2020prl}. \label{fig:contact}}
\end{figure}

According to Eq.~\eqref{eq:ejection}, the ejection spectrum within linear response contains the same set of eigenstates as the injection spectrum but with a prefactor that describes the initial occupation. Thus, at low temperatures, there is a strong suppression of the repulsive branch and only the attractive polaron appears in the ejection spectrum. The spectrum also contains information about the thermodynamic quantities for the interacting impurity-medium system, i.e., the relative free energy $\Delta \mathcal{F}$ features in Eq.~\eqref{eq:ejection}, while the Tan contact~\cite{Tan2008} (the conjugate variable to the scattering length)
\begin{align}
    C = \frac{8\pi m_r}{\hbar^2} 
    \frac{\partial \Delta\mathcal{F}}{\partial(-1/a)}
    \, ,
\end{align}
governs the high-frequency behaviour of the ejection spectrum: $\mathcal{I}_{\rm ej}(\omega) \to \frac{\sqrt{\hbar}}{4\pi^2\sqrt{2m_r}}\frac{C}{\omega^{3/2}}$~\cite{Schneider2010,Braaten2010}. At unitarity and $T=0$, the contact has the universal value $C=4.28k_F$ for the equal-mass Fermi polaron~\cite{Punk2009}, and this initially increases at finite temperature, displaying a marked non-monotonic behaviour, as shown in Fig.~\ref{fig:contact}. This is actually a signature of the impending polaron-molecule transition since it indicates that there are excited states with a larger contact $C$, i.e., an energy that is decreasing more rapidly with increasing interaction strength than the ground-state polaron~\cite{Liu2020prl}.

\subsection{Coherent dynamics}
It is also possible to investigate the coherent dynamics of impurities in the time domain, which can be ``ultrafast'' relative to the time scale set by the Fermi energy, $\tau_F \equiv \hbar/\ef$. A simple scenario is that of an interaction quench at zero temperature, where we start with the non-interacting ground state $\ket{\psi_0}$ and then switch on impurity-medium interactions at time $t=0$. To characterise the impurity dynamics, a natural quantity is the overlap function $S(t) \equiv \braket{\psi_0}{\psi(t)}$, which can be calculated from an expansion in energy eigenstates:
\begin{align}
    S(t) = \bra{\psi_0} e^{-i\hat{H}t/\hbar} \ket{\psi_0} = \sum_j |\!\braket{\psi_0}{\gamma_j}\!|^2 e^{-iE_jt/\hbar} \, .
\end{align}
In particular, it is easy to show that the Fourier transform is the injection spectrum:
\begin{align}
    \mathcal{I}_{\rm inj}(\omega) = \int^\infty_{-\infty} \frac{dt}{2\pi} e^{i\omega t} S(t) .
\end{align}
While $S(t)$ contains the same information as $\mathcal{I}_{\rm inj}(\omega)$, it gives access to both the phase and amplitude of the response, thus allowing one to probe the coherence between states.

\subsubsection{Ramsey interference} 
One such measurement that can probe coherent dynamics is Ramsey spectroscopy. Here one evolves an equal superposition of interacting $\ket{1}$ and non-interacting $\ket{2}$ impurity states in time, and then  interferes them to probe the amplitude and phase of $S(t) = |S(t)|\, e^{-i\phi(t)}$~\cite{Goold2011,Knap2012,Cetina2016}. 
For early times $t \ll \tau_F$, 
one can formally show that $S(t)$ is \emph{non-analytic} for a broad Feshbach resonance~\cite{Parish2016}\footnote{It is also non-analytic for a narrow Feshbach resonance, but at higher order in $t/\tau_F$~\cite{Cetina2016,Parish2016}.}
\begin{align}
    S(t) \simeq 1 - \frac{8e^{-i\pi/4}(m_\up/m_r)^{3/2}}{9\pi^{3/2}} \left(\frac{t}{\tau_F}\right)^{3/2} \, .
\end{align}
This is a consequence of the short-range interactions and is independent of the statistics of the medium, thus holding for both Bose and Fermi polarons, and for finite temperature. Indeed, this prediction was recently confirmed in a Bose-polaron experiment~\cite{Skou2021}.

\begin{figure}
\centering
\includegraphics[width=\textwidth]{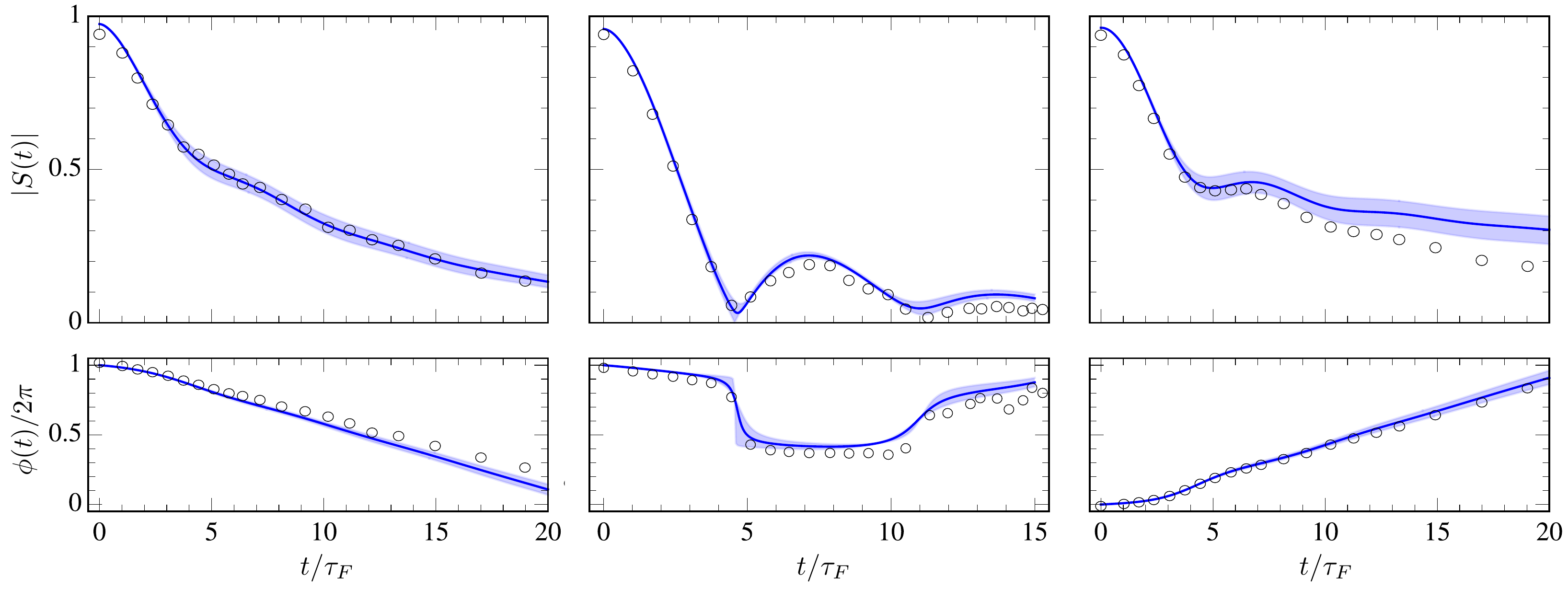}
\caption{Ramsey response $S(t)=|S(t)|e^{-i\phi(t)}$ of $^{40}$K impurities in a $^6$Li Fermi sea at $T/T_F\simeq 0.17$. From left to right, the results are shown for $1/\kf a=-0.86$, $1/\kf a=-0.08$, and for $1/\kf a=0.23$. Comparison between experiment~\cite{Cetina2016} and theory~\cite{Liu2019}. Adapted from~\cite{Liu2019}. \label{fig:ramsey}}
\end{figure}

At longer times, polarons form and the nature of the impurity becomes important. 
Figure~\ref{fig:ramsey} displays the Ramsey response for $^{40}$K impurities in a $^6$Li Fermi sea where we can clearly see the effect of the two polaron branches in Fig.~\ref{fig:rf} around unitarity. 
When the attractive or repulsive branch dominates for $1/\kf a \lesssim -1$ or $1/\kf a \gtrsim 1$, respectively, then the phase is determined by the corresponding polaron energy $E$, i.e., $\phi(t) = E t/\hbar$. Thus,  
the phase has  a negative slope in the attractive regime and a positive slope in the repulsive regime. However, close to unitarity, we observe phase jumps and oscillations of the amplitude due to the fact that we have a coherent superposition of both branches.

For an infinitely long-lived polaron, such as the ground-state attractive polaron, the long-time limit is $S(t) \to Z e^{-iEt/\hbar}$. However, for non-zero temperatures, the polaron generally has a finite lifetime and the amplitude eventually decays at long times $t\gg \tau_F$. 
This is due to a combination of thermal decoherence~\cite{Schmidt2018,Cetina2015} and ``many-body dephasing''~\cite{Adlong2020} which exists at $T=0$ for the repulsive polaron. Both are well described by the finite-temperature generalization of the variational approach with 1 particle-hole excitation (Section~\ref{sec:ansatz}), as evidenced by the excellent agreement with experiment in Fig.~\ref{fig:ramsey}.

\subsubsection{Rabi oscillations}
Another probe of coherent impurity dynamics is Rabi spectroscopy~\cite{Kohstall2012,Scazza2017,Oppong2019}, where the impurity states $\ket{1}$ and $\ket{2}$ are coupled via a continuous driving field  (see Fig.~\ref{fig:protocols}) 
resulting in oscillations between the two states. For a bare Rabi drive of frequency $\Omega_0$, 
the oscillation frequency $\Omega$ gives a measure of the polaron residue, i.e., $\Omega \simeq \sqrt{Z} \Omega_0$~\cite{Kohstall2012}, , which has enabled experiments to extract the residue for both attractive and repulsive polarons~\cite{Kohstall2012,Scazza2017,Oppong2019}. On the other hand, the damping of oscillations gives the (inverse) polaron lifetime~\cite{Adlong2020}.
Within the variational approach, the relevant observable is the population in each impurity state, e.g., the non-interacting state $\ket{2}$, which has the form~\cite{Parish2016,Adlong2020}
\begin{align}
    N_{\down,2}(t) = \Big|\sum_j |\alpha_{0,2}^{(j)}|^2 \, e^{-iE_jt/\hbar}\Big|^2 + \sum_{\k\q} \Big|\sum_j\alpha_{0,2}^{(j)^*}\alpha_{\k\q,2}^{(j)} \, e^{-iE_jt/\hbar} \Big|^2 \, ,
\end{align}
where the subscript 2 denotes the component in the $\ket{2}$ state and the energies $E_j$ (and associated eigenstates) now correspond to those for the Rabi-coupled system. The first term is connected to the injection spectrum of the Rabi-coupled system since it only contains the overlap with the non-interacting state. This is expected to dominate in the early times of the Rabi oscillations and it reveals the properties of the Fermi polaron stated above. The second term involves the dynamics of the dressing cloud which is not accessible in a standard linear-response measurement. As such, it is likely that this could provide further insight into the nature of the Fermi polaron.

\section{Outlook}
\label{sec:outlook}

The rapid progress in the understanding of the Fermi-polaron problem has been driven by the close interplay between theory and experiment, in part arising from the clean and highly tunable cold-atom system.
As discussed in these lecture notes, there is a remarkable agreement between theoretical predictions and experimental results for a range of key observables for the Fermi polaron, providing a prime example of precision many-body physics. Indeed, ultracold atomic gases can be viewed as an ideal ``quantum simulator'' for Fermi polarons, since the theoretical tools and ideas developed in this context are being translated to other systems, e.g., excitons in doped atomically thin semiconductors~\cite{Sidler2017}. Current open questions include the effect of medium interactions as well as the impact of longer ranged impurity-medium interactions such as dipolar interactions on the nature of the Fermi polaron.

An exciting current frontier is the role of correlations between Fermi polarons beyond the single-impurity limit. Here, even the sign of the polaron-polaron interactions appears to vary among different experiments~\cite{Tan2020,Fritsche2021,Muir2022} and the precise behaviour likely depends on whether the system is in or out of equilibrium and whether the impurities are thermal or degenerate. Ultimately, an understanding of such correlations could allow one to control polaronic quasiparticles and explore the quantum phase transitions associated with single-impurity transitions. 

Another important direction that may be realised in near-future experiments is the impact of exotic few-body bound states on polaron physics, both at the single-impurity level and at finite impurity density. For sufficiently large mass imbalance  ($m_\up/m_\down \gtrsim 8.2$ or $m_\down/m_\up \gtrsim 8.2$)~\cite{Kartavtsev_2007} there is the prospect of bound clusters (e.g., trimers) which are \emph{stable} unlike the Efimov bound states in the case of the Bose polaron~\cite{Yoshida2018}.
Experiments on $^{40}$K-$^6$Li mixtures~\cite{Jag2014ooa} have already revealed the first tantalising hints of the strong three-body correlations associated with these bound states, and these are expected to be even more important for mixtures with larger mass imbalance~\cite{Ciamei2022}. Furthermore, it would be exciting to investigate the effect of bound clusters and the associated few-body resonances on the dynamics of systems out of equilibrium, since this has implications for the fundamental question of how different quantum systems thermalise.

\acknowledgments
The authors are grateful to Kris van Houcke for sharing the data for Figs.~\ref{fig:polaron_results} and \ref{fig:polaron_results2}.
The authors acknowledge support from the Australian Research Council
Centre of Excellence in Future Low-Energy Electronics Technologies
(CE170100039).  JL and MMP are also supported through the Australian Research
Council Future Fellowships FT160100244 and FT200100619, respectively.

\bibliography{polaron2}

\end{document}